\documentclass[twocolumn,showpacs,preprintnumbers,amsmath,amssymb]{revtex4}

\usepackage{amsmath}
\usepackage{amssymb}
\usepackage{epsfig}
\usepackage{epsf,afterpage}
\usepackage{amssymb}
\newcommand{\be}{\begin{equation}}
\newcommand{\ee}{\end{equation}}
\newcommand{\lan}{\langle}
\newcommand{\rrr}{\rangle}

\usepackage{graphicx}
\usepackage{dcolumn}
\usepackage{bm}

\begin{document}

\title{Phenomenology of scale-dependent space-time dimension.}


\author{V.I.Shevchenko}
\affiliation{%
Institute of Theoretical and Experimental Physics
\\B.Cheremushkinskaya 25, 117218 Moscow, Russia
}%

\date{\today}

\begin{abstract}

Loop-mediated processes characterized by dynamical scale $M$
indirectly measure space-time dimension $d$ at this scale. Assuming
the latter to be scale-dependent and taking as examples
$B$-oscillations and muon $(g-2)$ experimental results we address
the question about constraints put by this data on $|4-d(L)|$ at
smaller distances, i.e. for $ML<1$. It is shown that sensitivity is
lost for $1/L$ around 300-400 GeV, and any value of $d(L)$ between 2
and 5 at this scale is compatible with the data.

\end{abstract}

\maketitle

\section{Introduction}

The main goal of modern high energy physics is to find and explore
phenomena beyond the Standard Model (SM). There are exciting New
Physics (NP) hints such as neutrino oscillations and dark matter.
 Nevertheless no experiment has shown so far direct and conclusive evidence in favor
of any particular NP scenario. Instead currently available data play
a role of constraints for different SM extensions. Despite some of
these constraints are rather tight, it is fair to say that many
reasonable TeV-scale NP scenarios discussed in the literature are
still far from being rejected.

 In a broad
phenomenological prospective almost all NP scenarios can be divided
into two large groups. The first one consists of the models which
extend particle content of the SM by adding some new particles,
according to this or that dynamical principle. The best known
example of this kind is supersymmetric extension of the SM. The
corresponding phenomenology is well studied, at least in case of
MSSM. Dynamical evolution of states in all models of this type takes
place on the standard Riemannian (3+1)-dimensional space-time
manifold of general relativity, whose internal dynamics is believed
to be governed by the genuine Planck scale $L_P = 1.6\times
10^{-35}$ m.

The second group of models suggests much more radical extension of
the SM. It is assumed that the picture of our familiar space-time as
smooth four-dimensional manifold is applicable only at low energies,
and becomes inadequate below some distance scale $L$ (which can be
much larger than $L_P$ and perhaps of (1-2 TeV)$^{-1}$ range). The
so called small extra dimensions and TeV-gravity scenarios
\cite{add1,add2,rs1,rs2} are well known examples of the theories of
this kind. The most attractive feature of these models is the
emergent nature of the Planck scale $L_P$.

Another line of reasoning, having its roots in seminal papers
\cite{wyler} and  \cite{mandelbrot}, is to some extent parallel to
extra dimensional picture. One can think of space-time geometry
becoming of discontinuous or of fractal type at small distances.
This idea has been explored from many different points of view:
conventional wisdom of Planck-scale quantum gravity fuzziness
\cite{wyler}, space-time foam models \cite{foam}, spin-networks in
loop quantum gravity \cite{loop}, dynamical triangulations
\cite{amb,ambjorn2}, fractal space-time structure in asymptotically
safe gravity \cite{reuter}, noncommutative geometry phenomenology
\cite{chai,noncomm}, modified commutation relations
\cite{dopli,kempf,das,ben}, modified dispersion relations in the
context of Finslerian geometry \cite{vis}, minimal length
phenomenology \cite{Hossenfelder1,Padmanabhan_base}, curved momentum
space \cite{kad} and other approaches. We refer the interested
reader to \cite{sd} for bibliographical review devoted to the models
of short distance space-time structure. In most of these approaches
one associates the corresponding NP length scale $L$ with the Planck
length $L_P$, despite nothing prevents to think about $L$ as being
different from $L_P$ (e.g., much larger).

The distinctive feature of all these extra approaches is the fact
that effective number of space-time dimensions felt by the
propagating particle depends on its energy/virtuality. One can think
of different signs of this dependence. While in versions of extra
dimensions scenarios which allow the SM fields to propagate in extra
dimensions they "open up" with the increase of particle momentum, in
fractal scenarios one can imagine low-dimensional dynamics in the
ultraviolet (UV) limit, and (3+1)-dimensional infrared description
as an emergent phenomenon.

Taking space-time dimension  $d$ as a free parameter it is
legitimate to ask a question about experimental measurement of this
quantity. In this way the quantity $\epsilon = (d-4)$ is to be
constrained by observations. There is extensive list of references
on the subject starting from \cite{ynd,jarls}. If $d$ is understood
as just a scale-independent constant devoid of any dynamics (as it
is done in the cited papers), the constraints come both from
celestial physics (Newton gravitational force law, perihelion
precession etc) and from atomic (Coulomb law, hydrogen spectra etc,
see e.g. \cite{yao}) as well as elementary particle physics (muon
$g-2$ etc). It is worth stressing that since any experiment actually
deals with finite energy-momentum/space-time resolution, one never
has access, strictly speaking, to the true UV dimension of a given
space-time and in this sense physical dimension always has to be
understood as resolution-dependent quantity.

Needless to remind that the idea of $d$ being different from 4 and
non-integer is very appealing from field theoretical point of view.
As is known since the advent of dimensional regularization
\cite{hooft,bol} and dimensional reduction \cite{siegel} methods,
the loop amplitudes which are divergent at integer values of $d$ (in
particular, for $d=4$) can be analytically continued and
self-consistently defined as finite quantities for noninteger values
of $d$. Usually understood as a convenient mathematical
regularization trick, this property may have a deeper meaning,
signalizing the preference for interacting quantum field theories to
live in non-integer dimensional world.  Moreover, despite it is
common practice in modern quantum field theory to understand the
regularization procedure as being formal, this is not the case for
some of the scenarios mentioned above, where space-time dynamics or
new particles can play a role of regulators for field-theoretical
amplitudes. For example, many of the NP scenarios with NP scale
$\Lambda$ can be described as a process-dependent change in loop
integration measure \be \int d^4 p \to \int d^4 p \> g(p^2,
\Lambda^2) \ee such that $g(p^2, \Lambda^2) \approx 1$ for $p^2 \ll
\Lambda^2$. On the other hand, at large $p^2$ the function $g(p^2,
\Lambda^2)$ is such that the integral becomes convergent. The best
known example of this pattern is given by cancelation of
divergencies in supersymmetric theories. As is well known minimal
length and alike scenarios also provide UV modification of the
measure (often process-independent) leading to convergence of loop
amplitudes.

In the present paper the question about phenomenological constraints
on space-time dimension at short scales is addressed, taking the
latter to be scale-dependent in a particular way.

 The concept of  scale-dependent dimension was thoroughly analyzed in
 studies of dynamics on multi-fractal sets (see, e.g \cite{kadanof}).
 There have been attempts to develop the corresponding formalism
 in field-theoretical context (see, e.g. \cite{eynk}).
In the present paper we explore some phenomenological consequences
of this picture being applied to experimental data on $B$ - mesons
oscillations and muon anomalous magnetic moment. The oscillations
phenomena are genuine quantum effects dominated by loop diagrams and
therefore they put into test, as any loop process does, the overall
integrity of quantum field theory. In particular, $K-\bar{K}$
oscillations constrain extensions of conventional quantum mechanics
caused by effects of nontrivial space-time dynamics
\cite{ellis2,mavr}. Speaking in a broad sense we are exploring the
same phenomenon (interplay between short-distance geometry and
quantum mechanics of mixing), but our concrete model is very
different from the one considered in \cite{ellis2,mavr} and
subsequent papers.

The organization of the paper is as follows. In the next Section II
we present our Ans\"atz and briefly discuss general features of the
corresponding phenomenology. In Section III the $B_s$ mixing rate is
analyzed and the exclusion plots of interest are given. We compare
the corresponding constraints with the ones provided by muon
anomalous magnetic moment data. The conclusions are presented in
Section IV. There is an Appendix in the paper, where relevant
formulas used in the main text are collected.

Since we are always interested in real parts of the amplitudes we
find it convenient to work in Euclidean metric and leave aside an
interesting question about Wick rotation in the context of
non-integer dimensional theories.

\section{The Model}

As is well known typical expression for amplitude in perturbative
quantum field theory has the following form: \be {\cal A} = {\cal
A}^{(0)}(q_i, m_i, e_i) + \sum\limits_{j=1}^{\infty}\> {\cal
A}^{(j)}(q_i, m_i, e_i)  \label{h} \ee where ${\cal A}^{(0)}(q_i,
m_i, e_i)$ is tree amplitude, $q_i, m_i, e_i$ stay for external
momenta, masses and couplings of interacting particles, while loop
amplitudes are given by \be {\cal A}^{(j)}(q_i, m_i, e_i) =
\left\{\prod\limits_{k=1}^{j} \int d^4 p_k \right\} \>{\cal
P}(p_k,q_i,m_i,e_i) \label{pole}\ee where $p_k$ are loop momenta
being integrated over and in the course of renormalization procedure
artificial dimensionful scale $\mu$ enters individual terms in
(\ref{pole}). In principle, many different strategies to continue
(\ref{h}) to $d\neq 4$ can be chosen. We adopt the picture of
dimensional reduction \cite{siegel}. The main feature which makes
this approach distinct from the standard dimensional regularization
is the following: one analytically continues in the number of
components $d$ of all loop momenta, but keeps the number of all
tensor and spinor field components fixed. Notice that all known
problems of dimensional regularization, notably $\gamma_5$-ambiguity
have their roots in the fact that nontrivial number of relevant
degrees of freedom is - contrary to the number of space-time
dimensions - an intrinsically integer quantity.

One immediate consequence of this prescription is that tree
amplitude ${\cal A}^{(0)}$ never gets modified (notice that this is
in contrast with many other approaches to non-integer dimensional
theories, e.g. \cite{ynd,yao}). All incoming and outgoing particles
by definition live in $d=4$, and the only affected element of
(\ref{pole}) is the loop integration measure.

From physical point of view there is some analogy with the weak
field expansion in gravity where one integrates the factor $\exp(-S)
= \exp(-\int d^4 z \sqrt{g}\> L[g_{\alpha\beta};\phi])$ over small
metric fluctuations $h_{\alpha\beta}(x)$ with respect to flat
background $\eta_{\alpha\beta} = g_{\alpha\beta}(x) -
h_{\alpha\beta}(x)$. In this case there is no question about
dimensionality of the bulk space, which always coincides with
$\eta_\alpha^\alpha$. Alternatively one can think of field theory
action defined on a fractal set $K$ (\cite{eynk}, see also \cite{sd}
and references therein): $S^{[K]} = \int_K d \mu_x \> L[\phi(x)])$
and a partition function $Z$  given by quantum average over
superposition of subsets $K$ of different and in general non-integer
dimensions $d$: $Z = \lan \exp(-S^{[K]}) \rrr_K $.

Then the crucial point is how such theory couples to external
currents. One has a choice - to take the coupling factor as $\exp
(i\int_K d\mu_x \> J(x)\phi(x))$ or to keep bulk expression $\exp
(i\int_{{\bf R}} d^n z \> J(z)\phi(z))$, where $K\subset {\bf R}$.
It is important to stress that these options would describe two
different kinds of physics. In the former case there is an
interference already of tree level processes mediated by particles
propagating over different $K$. One has to integrate over
$d$-dimensional external momenta and from phenomenological point of
view this brings us back to the constraints discussed earlier in
\cite{ynd,jarls,yao}. In terms of the theory of measurement this
situation corresponds to emitters and detectors being in coherent
superposition with the different states $K$. Another approach (the
one we actually explore) corresponds to the external currents always
living in $(3+1)$ dimensional world. External momenta are also
four-dimensional and tree processes receive no corrections. In other
words, macroscopic detectors are decohered with respect to the
space-time quantum fluctuations and hence detect only
four-dimensional particles, four-dimensional momenta etc. This
option seems to be physically more reasonable and we have chosen it
in the rest of the paper.

For phenomenological applications of our interest here we confine
attention to one-loop processes.  Introducing additional integration
over proper-time one can always rewrite the expression for ${\cal
A}^{(1)}$ as follows:
 \be
{\cal A}^{(1)} = \int\limits_0^\infty \frac{d\tau}{\tau} \tau^{-s}
\int d^4 p \>\exp(-\tau p^2)\> {\cal F}(\tau, q_i,m_i,e_i)
\label{e3} \ee where the parameter $s$ controls the degree of
divergence. In coordinate space this is replaced by corresponding
 expression for one-loop effective action
\be \Gamma[\phi] = \int\limits_0^\infty \frac{d\tau}{\tau} \> \int_K
d \mu_x \>\lan x | \exp\left(- \tau {\cal O}[\phi]\right) | x\rrr
\label{action} \ee where the operator $\cal O[\phi]$ encodes all
information about dynamics of the theory and Green's functions can
be extracted from (\ref{action}) by the standard technique. In case
of $K = {\bf R}^{4} $ we have conventional theory in flat
four-dimensional Euclidean space with the measure $\int d\mu_x =
\int d^4 x$. If the space $K$ is curved or has boundaries, one is to
apply Schwinger-De Witt technique \cite{dewitt} to get answer in
terms of expansion in powers of $\tau$ with the leading term given
by \be \lan x | \exp\left(\tau \Delta \right) |x \rrr \sim
\frac{1}{(4\pi \tau)^{\frac{d}{2}}}
 \label{eq214} \ee where $d=d(x)$ is
Hausdorff dimension of $K$ at the point $x$, $\Delta$ is
$d-$dimensional Euclidean Laplace operator, and the intrinsic
dimension of random walk on $K$ is taken to be equal to 2, as for
the standard Brownian motion (see detailed analysis in
\cite{mosco}). The value of $d=d(x)$ is defined by \be
\lim\limits_{r\to 0} \> \left(\frac{L}{r}\right)^{d}
\frac{\Gamma\left(\frac{d}{2}\right)}{2\pi^{d/2}}\> \int_{B_r}
d\mu_x = L^4 \label{eq02}\ee where $B_r \subset K$ is a ball of
radius $r$ centered at the point $x$ and $L$ - dimensionful scale,
characterizing the set $K$. Notice that (\ref{eq214}), (\ref{eq02})
are valid even for fractal set $K$ having no differentiable
Riemannian structure.

It is very interesting that the integration measure in
(\ref{action}) given for $K = {\bf R}^{4} $ by $\int d\tau /
\tau^{d/2+1} \int d^d x$ can be understood \cite{gopakumar} as
corresponding to curved space ${AdS}_{d+1}$ with the coordinates
$(x,\tau)$. If $h_{\mu\nu}(x)$ is a metric tensor on $K$, the
$d+1$-dimensional metric can be chosen as $ ds^2 = \tau^{-2} d\tau^2
+ \tau^{-1} h_{\mu\nu}(x,\tau)dx^{\mu}dx^{\nu} $ with the condition
$h_{\mu\nu}(x,\tau = 0) = h_{\mu\nu}(x)$ (notice that trivial choice
$h_{\mu\nu}(x,\tau) \equiv h_{\mu\nu}(x)$ would not be a solution of
Einstein's equations). Thus if one tries to keep the above
geometrical interpretation, the properties of the space $K$ become
$\tau$-dependent and factorization of the measure is lost, while the
Riemannian structure of the manifold is kept intact.

In the present paper we relax the latter condition and investigate a
particular deformation of (\ref{e3}) and (\ref{action})
corresponding to dimension of $K$ being $\tau$-dependent: $d\to
d(\tau)$. At the same time the space $K$ is taken as homogeneous in
the sense that $d$ does not depend on $x$. We assume the following
natural asymptotic conditions: $d(\infty)=4$, $d(0) = d$ where $d$
is some "true" ultraviolet space-time dimension. A typical scale
${\bar{\tau}}$ where the transition takes place is taken as a free
parameter of the model $L^2$.  The corresponding physics is outlined
in the introduction: at small virtuality, corresponding to $\tau$
much larger than $\bar{\tau} = L^2$,
 one has the standard 4-dimensional dynamics, $d(\infty)=4$.
 This condition is of prime importance for
the approach to preserve unitarity. It guarantees that virtual particles on-shell are indistinguishable from real ones and propagate in the same four-dimensional space-time. The situation is analogous to that in  dimensional reduction method (see discussion in \cite{siegel}).

Unfortunately we have no guiding physical principle to fix $d(\tau)$
dependence. One might appeal to numerical simulations from
\cite{amb,ambjorn2} or analytical results from recent papers
\cite{ben,horava}. However since our attitude here is mostly
phenomenological, we find it convenient to choose a particular
Ans\"atz for this function, which allows to proceed with analytical
computations. We have chosen the following one: \be \int \frac{d^4
p}{(2\pi)^4} \> \exp(-\tau p^2) \to \frac{1}{(4\pi
\tau)^{\frac{d(\tau)}{2}}} = \frac{f(w,d)}{(4\pi \tau)^2}
\label{function} \ee where $w=\tau / \bar\tau$ and the measure
deformation corresponds to $f(w,d) \neq 1$: \be f(w,d) = g(w) + (1-
g(w)) w^{2 -\frac{d}{2}} \label{function1} \ee The function $g(w)$
must obey obvious asymptotic conditions $g(0)=0$, $g(\infty) = 1$.
We have studied two particular choices for $g(w)$ making analytical
calculations possible: \be g_1(w)
=\left(1+\frac{1}{w}+\frac{1}{2w^2}\right)\exp(-1/w) \label{ch1} \ee
\be g_2(w) = 1- \frac{1}{\cosh(w)} \label{ch2} \ee The latter
function approaches $w=\infty$ asymptotic exponentially and $w=0$
polynomially, while the former one does it in the opposite way (the
chosen pre-exponential factor for $g_1(w)$ makes large-$w$
convergence faster). Certainly, we expect the results to depend on
gross features of the weight functions and not on the details how
they approach their asymptotic limits, and indeed this is what has
been found.

One can see that the function $f(w,d)$ switches between
4-dimensional dynamics in the infrared (large $w$) and
$d$-dimensional dynamics in the ultraviolet (small $w$). One can
interpret (\ref{function}) by saying that at typical virtuality
$1/\tau$ particles propagate as if effective dimension of space-time
would be \be d(w) = 4 - \frac{\log f^{2}(w,d)}{\log w} \ee The
character of this transition is controlled by the choice of $g(w)$.
We plot the functions $d_1(r)$, $d_2(r)$, $r\sim \sqrt{w}$
corresponding to two choices (\ref{ch1}),(\ref{ch2}) in Fig.1,
taking by way of example $d=2$. We have adjusted scales in such a
way that $L=L_2 = 2 L_1$, where $L_{1,2}$ correspond to the choices
$g_{1,2}(w)$. This allows to have roughly the same transition region
for two different functions (\ref{ch1}),(\ref{ch2}).  Notice that
for $d=4$ the function $d(\tau)\equiv 4 $ and does not depend on
$L^2$, since the scale $\bar\tau = L^2$ has been chosen as a scale
where dimensional reduction $4\to d$ happens.

At formal level phenomenology the expression (\ref{function}) leads  to
is in close correspondence with the one discussed in \cite{hill}.
Notice that we do not constrain $d$ to be larger than 4, or to be
integer with the increase of momentum scale, while the approach of
\cite{hill} adopts the extra dimensions logic, where effective
dimensionality of space-time always increases at small distances by integer steps.
Another, more technical thing is that our approach
preserves manifest $O(4)$ symmetry and gauge invariance. Indeed, the
general structure of one-loop effective action is such that the
integrand for the proper time integral is proportional to the trace
of the corresponding Wilson loop and is gauge-invariant by itself,
hence gauge invariance cannot be broken by any measure deformation $\int d\tau \to \int d\tau f(\tau)$.

Since the phenomenology of (\ref{function}) corresponds to changes
of ultraviolet behavior of the Green's functions, it poses a
question about validity of the whole approach in field theoretical
framework. This issue will be addressed elsewhere, here we only
notice that since deviation of $f(w)$ from unity parameterizes the
NP features as small corrections to the SM answers, we are actually
never in the regime where the effects caused by $d\neq 4$ become
dominant. Speaking differently, the loop integrals for SM
observables we look at are always saturated by the values of $\tau$
much larger than $L^2$.

 In the next section we apply the Ans\"atz (\ref{function}) to the
weak ($B$-oscillations)  and
electromagnetic (muon anomalous magnetic moment) loop processes.

\section{Applications}

Loop mediated processes have always played important role in
exploration of not yet discovered degrees of freedom. We are
interested to check the sensitivity of the selected electroweak and
electromagnetic observables to the UV measure deformation suggested
above. Recent experimental observation of oscillations of neutral
$B_s$ mesons complement the well known result for $B_d$
oscillations, and the current number is \cite{hfag} \be \Delta M_s =
(17.78 \pm 0.12) \> {\mbox{ps}}^{-1}\ee This result is consistent
with the SM expectations. For the models we are discussing there is
no principal difference between $B_d$ and $B_s$ cases. We consider
$\Delta M_s$ in what follows because theoretical uncertainty is
slightly smaller for this quantity.

It SM the mass difference between "heavy" and "light" mass
eigenstates, determining the oscillation frequency, is given by
\cite{wdbll}: \be \Delta M_s = \frac{G_F^2 M_W^2}{6\pi^2} (V^*_{ts}
V_{tb})^2 M_{B_s} {\hat\eta}^B {\hat{B}}_{B_s} f^2_{B_s} S_0(x_t)
\label{poi}\ee where the short-distance part (in effective
Hamiltonian sense) of this expression is represented by QCD
correction ${\hat\eta}^B = 0.552$ and the function $S_0(x_t)$, $x_t
= m_t^2/m_W^2$,  first computed in \cite{vys,il}, whose exact form
can be found in Appendix A. The status of this theoretical
prediction and space left for NP is reviewed in \cite{bf}. The
quantities entering (\ref{poi}) are measured or theoretically
computed with uncertainties, whose budget is conservatively
summarized in the table below:
\begin{center}
\begin{tabular}{|c|c|c|c|}
\hline  $\left[\frac{\delta \Delta M_s}{\Delta M_s}\right]_{exp}
$&$\left[ \frac{\delta (
f_{B_s}\sqrt{B_{B_s}})}{f_{B_s}\sqrt{B_{B_s}}}\right]_{lat}
$&$\left[\frac{\delta S_0(x)}{S_0(x)}\right]_{\delta m_t} $ &$
\left[\frac{\delta (V^*_{ts}V_{tb})^2}{(V^*_{ts}V_{tb})^2}\right]_{un}$\\
 \hline
0.7\%  & 8-12\%& 2.6\%& 2.0\%\\
 \hline
\end{tabular}
\end{center}
The uncertainties in $G_F$ and in the mass parameters are
negligible. Notice that to get independent prediction for $\Delta
M_s$ one has to consider CKM factors $(V^*_{ts}V_{tb})^2$ as input
parameters, to be determined by some other observables. The
uncertainty indicated above corresponds to unitarity-based
determination of $ |V^*_{ts}V_{tb}|_{un} = |V_{cb}| [ 1 -
\frac{\lambda^2}{2} (1- 2 R_b \cos\gamma)] $, and this relation is
supposedly unaffected by NP.

Needless to say that for comparison of theoretical predictions and
experimental data all uncertainties should be taken into account
together. Roughly speaking, NP contribution to the short-distance
function at the level of, say, 6\% would be washed out by large
uncertainty in soft hadronic part and in this sense not seen as a NP
signal, despite such deviation is larger than the theoretical
uncertainty in short distance function itself. It is known that the
theoretical uncertainties in the ratio $\Delta M_d / \Delta M_s$ are
smaller than in the numerator and the denominator separately, but in
minimal flavor violating scenario discussed by us here this ratio
provides no information about NP since the short distance functions
exactly cancel in it. The crucial importance to reduce theoretical
uncertainties by simultaneous consideration of different hadronic
inputs and not by cancelations of this kind is stressed in another
context in \cite{we}.

The prescription (\ref{function}), (\ref{function1})
 allows to compute new short-distance function in
analytic form. We chose the parameter $x_t={\bar{m}}_t^2/m_W^2=4.20$
corresponding to ${\bar{m}}_t(m_t) = (164.7\pm 2.8)$ GeV \cite{hfag}
and define the normalized ratio \be s(L,d) = S(L,d)/S_0 \ee where
$S(L,d)$ is the new short-distance function and the SM
short-distance function $S_0\equiv S_0(x_t)$ is given in Appendix A.
One can check that \be s(L,4) \equiv 1 \;\; \mbox{and} \;\;
\lim\limits_{L\to 0} s(L,d) = 1 \ee as it should be.

The Figure 2 represents the results as contour plots of $s(L,d)$ for
the choice (\ref{ch1}).  Dashed contours correspond to $\pm 10\%$
deviation of $S(L,d)$ from the SM result $S_0$, thin contours - to
$\pm 5\% $ deviation and thick contours - to $s(L,d)\equiv 1$. The
results for the choice (\ref{ch2}) are depicted at Figure 3.
Sensitivity to the scale $L$ is lost for $d=4$, hence the horizontal
line $s(L,4)=1$ on Figs.2,3. The geography of allowed and excluded
regions at the corresponding level of accuracy is clear from the
pictures. We see good level of qualitative similarity between the
figures, illustrating the expected robustness of the answer with
respect to asymptotic properties of chosen transition weights
$g_1(w), g_2(w)$.

It is interesting to see that at the scale $L^{-1}$ as small as
300-400 GeV the experimental data on oscillations put almost no any
constraints on dimensionality of space-time $d$ at this scale.
 It is
instructive to compare this result with the $R^{-1} > 600$ GeV bound
on minimal universal extra dimension radius $R$ from $B\to X_s
\gamma$ data recently obtained in \cite{haisch}.

It is interesting to reconcile the above results with the
constraints coming from non-flavor physics, namely from QED. Of
prime interest in this respect is the anomalous muon magnetic
moment. The reader is referred to the reviews \cite{knecht,recer},
recent updates \cite{hagi,hagi2,expmu} and references therein for
introduction into the subject. Current experimental value for \be
a_\mu = \frac12 (g_\mu -2) = F^{SM}_2(k^2 = 0) \ee is given by (see
\cite{recer} and references therein): \be a_\mu =
11659208.0(6.3)\times 10^{-10}
 \label{exp} \ee
 There is strong interest on this subject since some $\sim 3 \sigma$ discrepancy between theoretical SM
prediction for this quantity and the experimental result (\ref{exp})
is seen. The former, however, suffers from hadronic uncertainties at
the level about 0.5 ppm, i.e. comparable with experimental accuracy,
therefore it is very difficult if not impossible to deduce solid
positive statement about the meaning of such discrepancy. But
regardless the status of the SM theoretical prediction it is obvious
that no NP scenario may bring corrections to the SM which are in
contradiction with the experimental result (\ref{exp}). The
unprecedented accuracy of current experimental value $\delta a_\mu /
a_\mu = 0.54$ ppm - makes this requirement especially challenging.
It is clear, that in our case the dominant correction comes from the
one-loop term, which is proportional to $\alpha$. Thus we must
require the correction already at the first loop to be smaller than
combined experimental uncertainty $\delta a_\mu = 6.3 \times
10^{-10}$: \be w(L,d) = \frac{\left| F_2(L,d) - F_2^{SM} \right|
}{\delta a_\mu } \lesssim 1\ee where $F_2(L,d)$ is the new
short-distance function, while the SM answer is given at one loop by
the classic Schwinger result $F_2^{SM} = \alpha/(2\pi)$. The
corresponding Feynman integral can be found in Appendix A. The
results are presented on Figs.4,5. We observe again expected
similarity between the figures.  It is seen that despite
unprecedented experimental accuracy the muon anomalous magnetic
moment data cannot be competitive to electroweak loop observables in
the discussed respect.

\section{Conclusion}

We have studied some elements of phenomenology of scale-dependent
space-time dimension. The chosen prescription corresponds to
invariant deformation of the integration measure in loop integrals.
The virtual particles propagates in $4$ dimensions at typical
distances larger than $L$ and in $d$ dimensions at typical distances
smaller than $L$, thus measuring space-time dimension at the scale
$L$ (Figure 1).

The results in form of the contour plots are shown on Figs.2-5. They
can be interpreted as indirect measurements of space-time dimension
at the scale $L$ by experimental data on $B_s$ oscillations and muon
$(g-2)$. Since typical process characterized by dynamical scale $M$
indirectly measures space-time dimension $d$ at this scale we could
not expect good sensitivity of our observables to $d(L)$ for $ML\ll
1$ if not for special reasons. In case of muon $(g-2)$ data the
reason obviously is unprecedented accuracy of experimental result,
which partly compensates smallness of $m_\mu L$. Anyway we find it
remarkable that roughly speaking any number of dimensions between
$5$ and $2$ at a scale as small as 350 GeV is compatible with
experimental data. This is to be compared with $10^{-7} - 10^{-9}$
bounds on $\epsilon = |d-4|$ from \cite{ynd,jarls}.

 It a sense, this is rather general situation, as can
be seen comparing our results with that of
\cite{Buras_extra1,Buras_extra2,BMRS1,BRS2,HHBS,klink}. Having just
two-dimensional minimal flavor violating NP parameter space is
enough to open up possibilities for rather low energy scale of New
Physics.

\acknowledgments

The author acknowledges discussions with R.Fleischer and V.Orlovsky.
The work is supported by the INTAS-CERN fellowship 06-1000014-6576
and partly by the grant for support of scientific schools NS-4961.2008.2. and
RFBR Grant 08-02-91008.

\appendix

\section{Appendix A}

To be self-contained we collect here explicit formulas used in the
main text to compute the contour plots of interest.

1. The basic integrals are given by \be I(s,d,y,\bar\tau)=
\int\limits_{0}^{\infty} \frac{d\tau}{\tau^s} \cdot
f(\tau/{\bar{\tau}},d) \exp(-y\tau) \ee and can be expressed in
terms of Bessel functions of the second kind and $\Gamma$-functions
in case of $f(\tau/{\bar{\tau}},d)=f_{1}(w,d)$ and it terms of
$\Gamma$ and $\zeta$-functions for
$f(\tau/{\bar{\tau}},d)=f_{2}(w,d)$. The full expressions are rather
cumbersome.

2. The short-distance function $S_0=S_0^{WW} + S_0^{WH} + S_0^{HH}$
is given by $$ S_0^{WW} = 16\pi^2  m_W^2 \int \frac{d^4 p}{(2\pi)^4}
\frac{m_t^4}{(p^2 + m_W^2)^2}\frac{1}{p^2}\frac{1}{(p^2+m_t^2)^2}
$$
$$
 S_0^{HH} =
4\pi^2 \frac{m_t^4}{m_W^2} \int \frac{d^4 p}{(2\pi)^4} \frac{1}{(p^2
+ m_W^2)^2}\frac{p^2}{(p^2+m_t^2)^2} $$ $$ S_0^{HW} = 32\pi^2 m_t^4
\int \frac{d^4 p}{(2\pi)^4} \frac{1}{(p^2 +
m_W^2)^2}\frac{1}{(p^2+m_t^2)^2} $$  with the result \be S_0 =
\frac{4x_t - 11 x_t^2 +x_t^3}{4(x_t -1)^2} + \frac{3x_t^3\log
x_t}{2(x_t -1)^3} \ee where $x_t = m_t^2 / m_W^2$.

3. The SM one-loop contribution to the muon anomalous magnetic
moment is given by the following Feynman integral \be F_2^{SM} =
8e^2 \int\limits_0^1 x^2(1-x) \int \frac{d^4 \tilde{p}}{(2\pi)^4}
\frac{1}{(\tilde{p}^2 + x^2)^3} \ee The function $F_2(L,d)$
corresponds to the replacements (\ref{function}) in this integral,
with the proper account of the loop momentum rescaling $p=m_\mu
\tilde{p}$.

\newpage

\begin{figure}
\epsfxsize=6cm 
\epsfbox{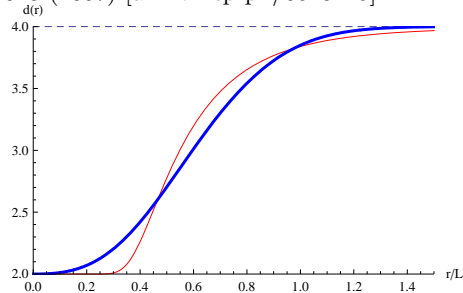}
\caption{Scale-dependent space-time dimension as
a function of distance $r$ in units of $L$, thin (red) line
corresponds to the choice (\ref{ch1}), thick (blue) line corresponds
to the choice (\ref{ch2}). }
\end{figure}
\begin{figure}
\epsfxsize=6cm
\epsfbox{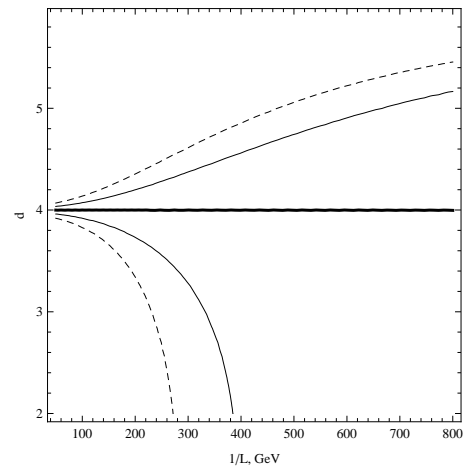}
\caption{Contour plot for the function  $s(L,d)$ with the choice (\ref{ch1}).
Contours correspond to $s(L,d)=1\pm 0.1$ (dashed), $s(L,d)=1\pm
0.05$ (thin) and $s(L,d)= 1 $ (thick).}
\end{figure}
\begin{figure}
\epsfxsize=6cm
\epsfbox{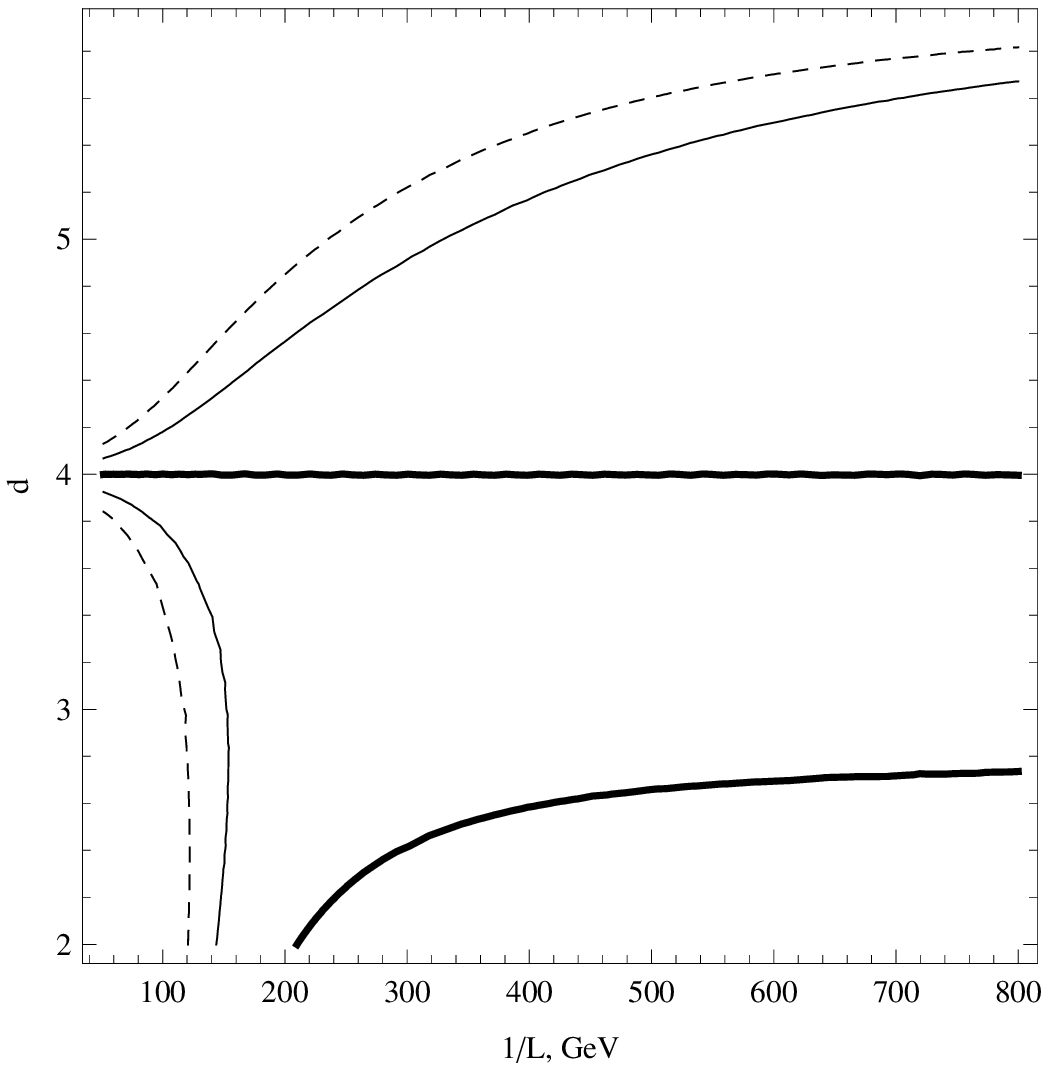} \caption{The same as Fig.2, for the choice
(\ref{ch2}).}
\end{figure}


\begin{figure}
\epsfxsize=6cm \epsfbox{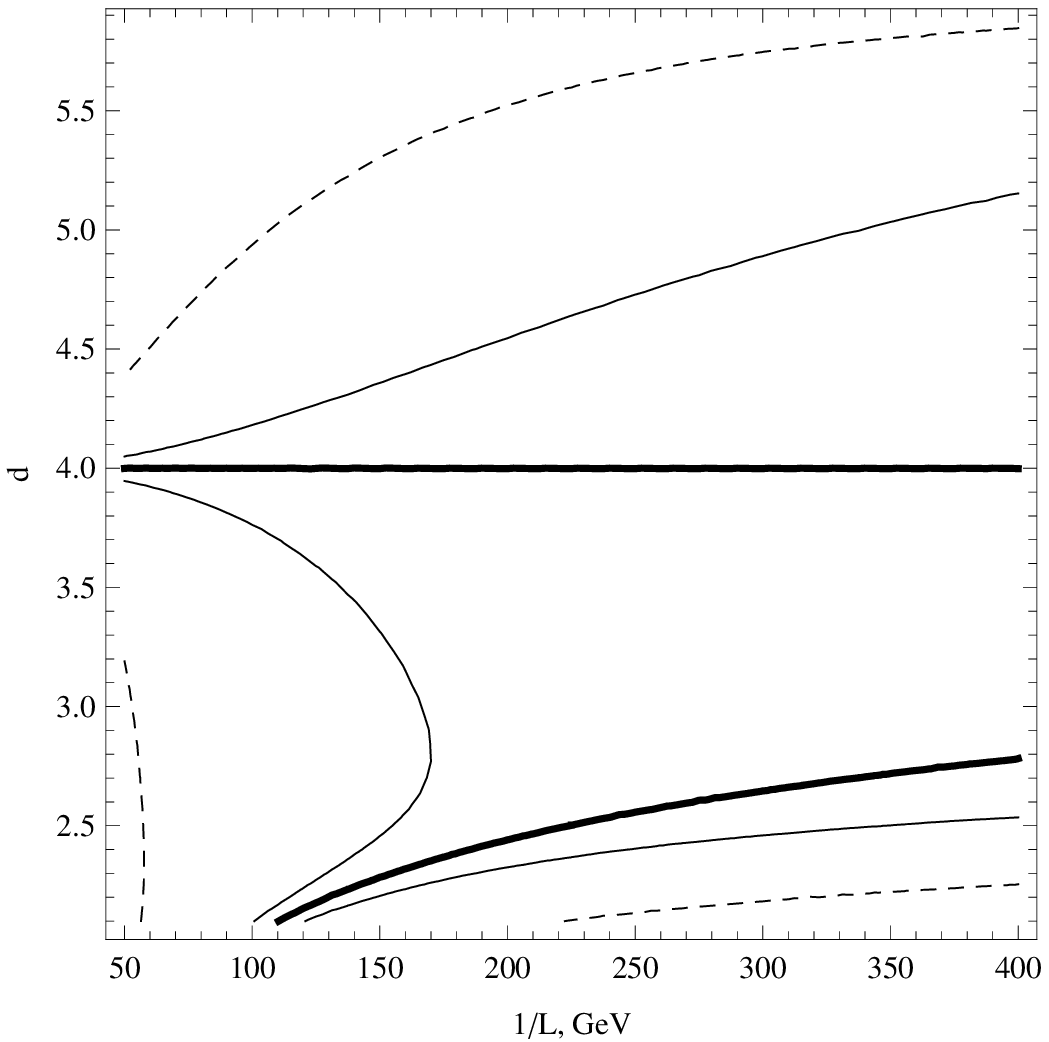} \caption{Contour plot for the
function $w(L,d)$ with the choice (\ref{ch1}). Contours correspond
to $w(L,d)=\pm 1$ (dashed), $w(L,d)=\pm 0.1$ (thin) and $w(L,d)= 0$
(thick). }
\end{figure}\begin{figure}
\epsfxsize=6cm  \epsfbox{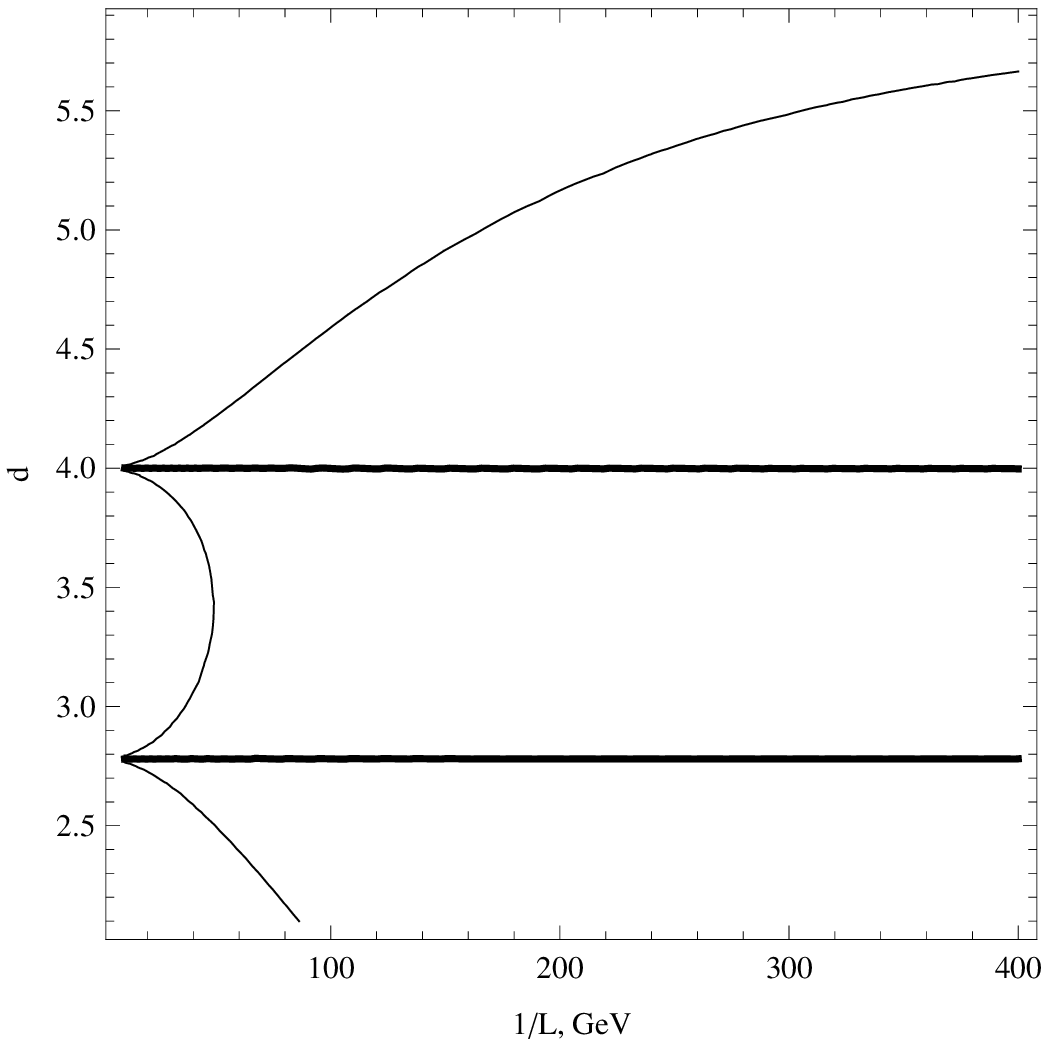} \caption{The same as Fig.4, for
the choice (\ref{ch2}).  }
\end{figure}

\end{document}